# Prospects for giant optical nonlinearity on a chip

J. Hwang and E. A. Hinds

*Centre for Cold Matter, Blackett Laboratory, Imperial College London,
Prince Consort Road, London SW7 2AZ*

We point out that individual organic dye molecules, deposited close to optical waveguides on a photonic chip, can act as single photon sources and can also provide localised, giant optical nonlinearities. This new atom-photon interface may be used as a resource for processing quantum information.

INTRODUCTION

Single photons are exceedingly attractive as a basis for quantum information processing (QIP) because they are robust against dephasing and are able to encode information in several degrees of freedom such as polarization, time bin and path.[1] The main challenge for photonic QIP is to implement efficient interactions *between* photons. Since ordinary materials are not sufficiently nonlinear to achieve this, it was initially believed that sophisticated methods such as electromagnetically induced transparency[2] or cavity QED[3] would be essential.[4] In 2001, however, Knill, Laflamme, and Milburn suggested a fundamentally different approach, known as linear optical quantum computing (LOQC).[5] They proved it is possible to create a universal quantum computer with linear optics alone, using Hong-Ou-Mandel (HOM) interference[6], feed-forward and photodetection. Two of these requirements have recently been realized on a photonic chip by the group of O'Brien.[7] Optical waveguides on the chip provide networks of interconnected interferometers with high visibility and with excellent control over the alignment and purity of the optical elements. These chips promise a robust platform for quantum logic that is scalable in principle, but only at the cost of prohibitively large resources. By introducing a nonlinear element into the chip dramatic savings in the required resources can be obtained. As a simple, practical way to achieve this, one might want to use single organic dye molecules in a transparent organic matrix as the nonlinear material. At cryogenic temperature, each molecule behaves as a 2-level quantum system and has a strong nonlinear interaction with passing photons, provided these are confined within an area comparable to the resonant interaction cross section.[8,9] High-index dielectric waveguides can provide such small mode-field areas. To create a 2-dimensional network of nonlinearities, single organic dye molecules may be deposited on top of pre-fabricated networks of waveguides. The molecules are expected to induce substantial nonlinear phase shifts in the light. For a realistic geometry that we discuss below, the phase shift of one photon changes by more than a degree when a second photon is present. This strong nonlinearity offers a possible way to avoid the overhead of LOQC on a chip.

To realize the full potential of photonic QIP, sources and detectors must be integrated into the chip in a scalable way. Many groups are working to develop suitable detectors, see [10] for example. By contrast, the single photons used for QIP on a chip have so far been generated externally using parametric down-conversion, which does not seem to be a scalable approach. Semiconductor III-V quantum dots have promise as single photon sources on a chip, however, it remains challenging to extract photons efficiently from the high index semiconductor material, and to scale up to a large number of emitters.[11,12] Alternatively, trapped atoms or ions could also act as single photon sources, but these atomic systems have not yet emerged as a practical way to implement quantum logic on an optical chip because the complex functions of laser cooling and atom trapping are not easily integrated into the chip. By contrast, individual dye molecules offer a simple way to integrate several single-photon sources into an optical chip. When a short laser pulse excites one or more of these molecules, each molecule emits a single photon into the nearby waveguide, thereby acting as a source of transform-limited single photons with high fidelity and high repetition rate.[13,14] The method we suggest here is considerably simpler than existing schemes and is scalable. The waveguides are fabricated first, the molecules are then deposited on top, and finally those of interest are selectively addressed and coupled to the waveguides.

## I. SINGLE MOLECULES AS QUANTUM OPTICAL DIPOLES

A convenient operating wavelength might lie in the wavelength range of 780 – 900 nm, where silicon photodiodes have high quantum efficiency and high-index waveguide materials such as SiN and GaP have good transparency. A suitable molecule for this purpose is dibenzoterrylene (DBT) operating in the vicinity of 785 nm, whose structure is illustrated in Fig. 1(a). The photophysical properties of DBT have been thoroughly studied in an anthracene matrix.[15,16] It is more convenient to work with matrices that are liquid at room temperature, such as n-hexadecane[17] or methyl methacrylate (MMA)[18] so that the molecules can be introduced onto the photonic chip in droplets or spin coated. These matrix molecules are also shown in Fig. 1(a). On insertion into the cryostat, the matrix material crystallizes to form a Sh'polskii matrix,[19] moulded to the shape of the nanoscale surface structures. At temperatures below 2 K, the DBT molecule acts as simple two-level system with a strong electric dipole transition. Each molecule can be individually addressed and remains trapped indefinitely in the solidified solvent.

Figure 1(b) shows the level scheme of a single DBT molecule in the host matrix. The zero-phonon-line (ZPL) of the molecule connects the ground state ($S_{0,v=0}$) to the ground state of the first electronically excited manifold ($S_{1,v=0}$). This forms the 2-level quantum system of interest, with a lifetime-limited linewidth of 30 MHz. Since the non-radiative decay from the $S_{1,v=0}$ state is negligible, the molecule can be made to emit a single photon simply by exciting it with a short laser pulse.[13] Excitation on the $(0-1)$ transition, shown dash-dotted in Fig. 1(b), is followed by fast vibrational relaxation to $S_{1,v=0}$ and subsequent decay to $S_{0,v=0}$. This conveniently separates the excitation frequency from the frequency of the decay photons. A fraction of the spontaneous decays are red-shifted Raman lines as shown dashed in Fig. 1(b). Because the vibrationally excited $S_{0,v\neq0}$ states relax rapidly to $S_{0,v=0}$, these lines are broad and not useful for our purpose. The branching ratio for emitting a ZPL photon varies from one DBT molecule to another over the range 0.1 – 0.5, depending on the local environment. There is also some inhomogeneous variation of the ZPL frequency. Thus, although a given molecule emits identical photons into the waveguide, these photons may not interfere with photons from another molecule. This inhomogeneous shift can be removed by applying a local electric field to each molecule that Stark-shifts them to a common resonance frequency, as was recently demonstrated with two molecules.[20] Individual molecules, tuned to have identical transition frequencies and coupled to waveguides, can serve as integrated photon sources for quantum information processing on an optical chip.

Molecules that have a strong electric dipole transition can also be used as sources of optical nonlinearity on chip. It was recently demonstrated that a single DBATT molecule embedded in n-hexadecane can obscure a substantial part of the light in a beam whose cross section is comparable to the scattering cross section of the molecule.[21] In such a beam, it was shown that the molecule behaves as a two-level quantum emitter whose nonlinear response is appreciable even at very low light intensity. This was seen through the appearance of Mollow sidebands in the fluorescence spectrum of the molecule.[21] In another proof of the large dipolar coupling, Rabi flopping of the two-level molecule could be observed in weak light pulses, where only a few hundred photons were enough to generate a π-pulse.[22] Most recently, it was shown that the single molecule can act as an absorber whose absorption coefficient can be manipulated by a control laser and can also be turned into gain when the population of the molecule is inverted.[23]

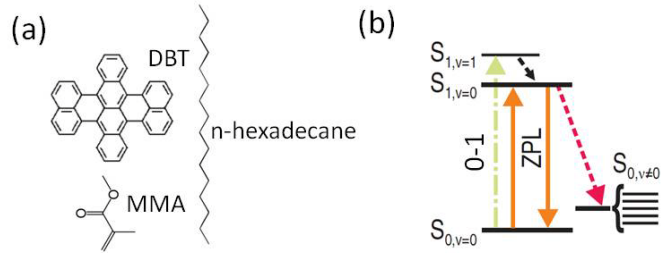

**Fig 1.** (a) Molecular structures of dopant molecule DBT and of matrix materials MMA and n-hexadecane. (b) Level scheme of DBT transition at 785 nm. Dash dotted arrow: excitation to produce population inversion. Solid arrow: narrow zero-phonon line transition.. Dashed arrow: broad Raman sidebands.

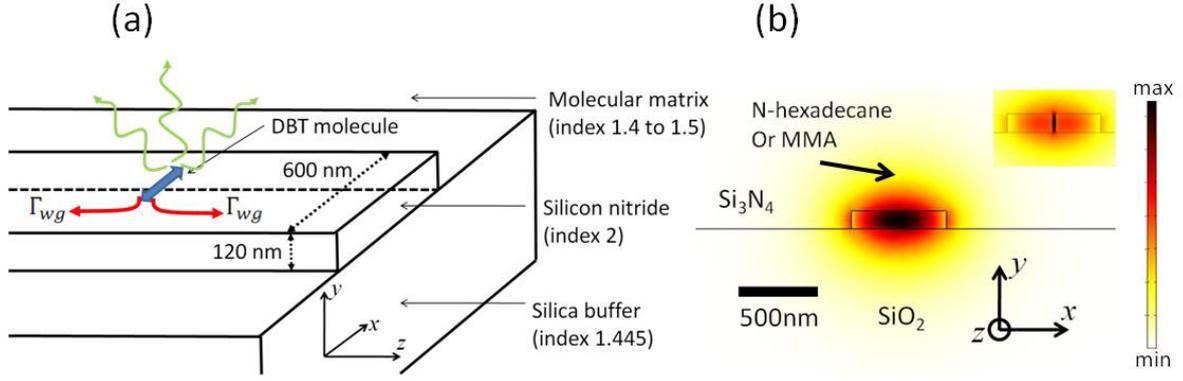

**Fig. 2.** (a) Example of waveguide design and illustration of molecular emission channels. A 120 nm-thick $Si_3N_4$ core (refractive index $n = 2$) is deposited on $SiO_2$ ($n = 1.445$) and covered with a molecular matrix such as n-hexadecane ($n = 1.434$) or MMA ($n = 1.42$). A molecule on the top surface of the core emits into the waveguide mode at a rate $\Gamma_{wg}$ in each direction. It also emits into the rest of the space. (b) Calculated transverse mode for such a strip waveguide, covered by n-hexadecane and operating at the ZPL frequency of the DBT molecule ($f/c = 785$ nm). Propagation is into page and colour indicates the magnitude of the electric field along *x* in quasi TE operation. MMA gives a similar result. *Inset*: the same waveguide but with a 40 nm gap filled with n-hexadecane.

In comparison with other solid state emitters such as nanocrystals containing diamond N-V centres[24], colloidal quantum dots[25], and self-assembled quantum dots[26], the very small size of the organic dye molecules offers a significant advantage. At practical concentrations, there can be several hundred molecules in an illuminated volume of 1 $\mu m^3$, compared with only a few for these other emitters. This greatly increases the probability of finding a molecule exactly where it is needed. Since the frequency of the ZPL is inhomogeneously distributed over ~1 THz, while the natural width is only about 30 MHz, there is no difficulty identifying the individual molecules within this volume by their laser-induced fluorescence, using simple far-field optics and a tuneable, narrow-band laser. One is almost bound to find a suitable molecule closely coupled to the waveguide at any desired position along its length. At each position where a molecule is needed, the resonance frequency of the selected molecule can be shifted to the desired photon frequency by the Stark tuning using small electrodes deposited on each side of the waveguide.

## II.   DESIGN OF WAVEGUIDE FOR LARGE COUPLING TO MOLECULES

If the cross sectional area of a light beam is small, the electric field of one photon is correspondingly large. Efficient molecule-waveguide coupling therefore requires a waveguide designed to have as small a mode cross section as possible. A molecule positioned suitably close to such a waveguide may be so strongly driven that one photon saturates the molecular polarisability, leading to an effective interaction between photons. Large dipolar coupling also leads to preferential emission of photons into the waveguide mode.

Figure 2(a) shows an example of a suitable waveguide design. The waveguide is a high index rectangular strip on a lower-index substrate. We consider Silicon Nitride ($Si_3N_4$), with a refractive index of 2, on silica with refractive index 1.45. The molecules dissolved in their molecular matrix are deposited on top. After cooling in a cryostat, the solvent forms a top layer of very similar index to the silica: 1.434 for n-hexadecane and 1.42 for methyl methacrylate. The $Si_3N_4$ layer is readily deposited on silica using Plasma-Enhanced Chemical Vapour Deposition (PECVD), which allows the thickness to be controlled accurately. At the 785 nm wavelength of DBT, we calculate that a thickness of 120 nm maximizes the evanescent field per photon at the position $\vec{r_0}$ of a molecule, taken as 20nm above the strip. We choose a width of 600 nm to ensure single-mode operation. Figure 2(b) shows the calculated mode distribution when the top layer is n-hexadecane. Using this distribution, we calculate that the effective mode cross section $A_{eff}$ for coupling to the molecule is

$$A_{eff} = \frac{\iint \varepsilon(\vec{r})|\vec{E}(\vec{r})|^2 dA}{\varepsilon(\vec{r_0})|\vec{E}(\vec{r_0})|^2} = 0.42\,\lambda^2, \tag{1}$$

$\varepsilon(\vec{r})$ being the relative permittivity at position $\vec{r}$ and $\lambda$ the transition wavelength in free space. This indicates that large coupling is possible between the molecule and the waveguide, since $A_{eff}$ is comparable with the free-space optical cross section for the molecular ZPL transition, $\sigma_{ZPL} = \frac{3\lambda^2}{2\pi}$.

Even larger coupling may be achieved by cutting a narrow slot along the waveguide in the propagation direction, as shown in the inset into Figure 2(b) for a gap of 40 nm. With such a narrow gap, our simulation shows that the propagating mode is largely unaffected, while the electric field in the gap is enhanced by the ratio of guide and slot permittivities.[27] Thus, the effective mode area for coupling to a molecule in this slot is only $0.10\,\lambda^2$.

## III.   PHOTON EMISSION INTO THE WAVEGUIDE

Figure 2(a) illustrates an excited DBT molecule placed 20 nm above the integrated waveguide. When the DBT molecule in n-hexadecane or MMA is excited at 785 nm, its decay is almost entirely radiative. Photons are radiated as travelling waves in the waveguide at a rate $\Gamma_{WG}$ for in each direction. There is also radiation into the rest of space.

Let us first consider the rate of radiation into the waveguide using Fermi's golden rule

$$\Gamma(\vec{r},\omega) = 2\pi|g(\vec{r},\omega)|^2 D(\omega) \tag{2}$$

where $g(\vec{r},\omega)$ is the dipolar coupling constant per photon for an emitter at position $\vec{r}$ with transition frequency $\omega$ and $D(\omega)$ is the photonic density of states. The waveguide supports only one transverse mode at the transition frequency, so the density of waveguide modes is found by counting longitudinal modes.[28,29] For waves travelling in one direction with a periodic boundary condition over a large length $L$, $D(\omega) = \frac{L}{2\pi v_g}$, where $v_g$ is the group velocity in the guide. For the same travelling wave mode, the rotating-wave dipole coupling is given by $|g|^2 = \frac{d_x^2\,\omega}{2\hbar\,\varepsilon_0 n^2\,L\,A_{eff}}$ when we treat the substrate, coated by matrix material, as a homogeneous dielectric of real permittivity $\varepsilon = n^2$ at the transition frequency of the molecule. Here $d_x$ is the dipole transition matrix element that couples to the *x*-polarised waveguide mode and $A_{eff}$ is defined in Eq. (1). As pointed out by Barnett et al.,[30] an additional factor of $\left(\frac{n^2+2}{3}\right)^2$ is introduced by the local field correction in the dielectric. Collecting these terms together we arrive at the spontaneous emission rate of travelling wave photons in one direction in the guide.

$$\Gamma_{WG} = \frac{d_x^2\,\omega}{2\,\hbar\,n^2\varepsilon_0\,A_{eff}\,v_g}\left(\frac{n^2+2}{3}\right)^2 \qquad (3)$$

It is convenient to compare this with the radiation rate in the matrix material without any waveguide, given by[30]

$$\Gamma_{rad} = \frac{d^2\omega^3}{3\pi\hbar\epsilon_0 c^3}\,n\left(\frac{n^2+2}{3}\right)^2. \qquad (4)$$

The first factor is the usual formula for radiative decay into free space, with $d^2 = d_x^2 + d_y^2 + d_z^2$. The effect of the dielectric is contained in the last two factors. The ratio of these two radiation rates is

$$\frac{\Gamma_{wg}}{\Gamma_{rad}} = \frac{1}{4}\frac{d_x^2}{d^2}\frac{\left(\frac{3\lambda^2}{2\pi\,n^2}\right)}{A_{eff}}\frac{(c/n)}{v_g} \approx \frac{1}{4}\frac{\left(\frac{3\lambda^2}{2\pi\,n^2}\right)}{A_{eff}}. \qquad (5)$$

The anisotropy of the molecule concentrates most of its transition moment along a particular direction. Therefore, when the molecule is suitably oriented with respect to the waveguide, $d_x^2/d^2 \approx 1$. For the thin waveguides we are considering here, much of the field mode is in the matrix material, of refractive index $n$, and this makes $v_g \approx c/n$, except when the frequency is close to the waveguide cutoff. Equation (5) gives us some insight into the requirements for creating photons efficiently in the waveguide. Specifically, the molecule should be aligned with respect to the waveguide, the effective mode area $A_{eff}$ should be small, and the group velocity $v_g$ should be low. For the waveguide considered in section II, Eq.(5) gives $\Gamma_{wg}/\Gamma_{rad} = 0.14$. This expression does not directly determine the ratio of photons in the guide to photons in the substrate/matrix because the rate for the latter is not equal to $\Gamma_{rad}$: it is affected by the presence of the guide.

In order to address this point, we simulated the problem numerically, using the MEEP programme[31] to calculate the power radiated by a classical dipole in a dielectric medium, both with and without the waveguide. The result with the waveguide, shown in Fig. 3(a), confirms that the power in each guided direction is indeed 14% of the power radiated into a homogeneous dielectric with no waveguide. We also find that the total radiated power remains almost unchanged (there is a 5% increase) when the waveguide is present. This indicates that the power coupled into the guide is diverted from power that would otherwise be part of $\Gamma_{rad}$, indeed we see explicitly that the angular distribution of the radiation in the bulk is strongly modified by the presence of the guide. We conclude therefore that 28% of the photons emitted are coupled into the waveguide, 14% in each direction.

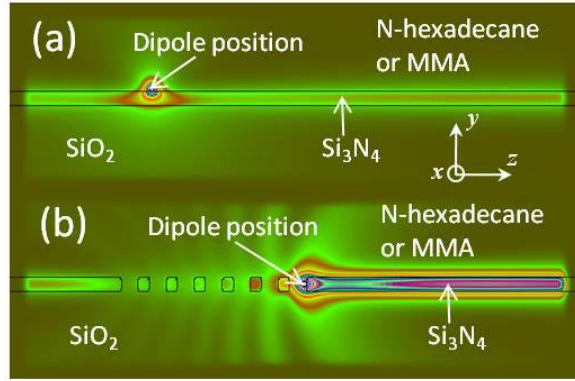

In order to deliver all these photons to the target, we consider placing a mirror on the left side of the molecule, so that the left-going field is reflected and adds constructively to the right-going field. With the molecule positioned at the antinode of the interference pattern, the coupling to the travelling output should increase from $\Gamma_{WG}$ to $4\Gamma_{WG}$, producing a rate

**Fig. 3** (a) Finite-difference time-domain simulation of dipole radiation coupled to the waveguide. The propagation axis is horizontal in the page. The dipole, placed 20 nm above the strip, is pointing out of the page (x-direction) exciting the quasi TE mode. (b) Uni-directional launching of single molecule emission using a Bragg reflector: a quarter wave stack of n-hexadecane blocks alternating with the $Si_3N_4$ blocks. The molecule is placed inside the right-most trench, 20nm from the end face of the strip.

that is 56% of $\Gamma_{rad}$. The corresponding numerical simulation, shown in Fig. 3(b), demonstrates that this idea works when we make a Bragg reflector by modulating the waveguide. Over 50% of the radiation is now coupled into the right-going waveguide and again the total radiation rate is well approximated by $\Gamma_{rad}$. We find that the Bragg mirror scatters some of the power into the matrix material, but this is not a large loss. Further simulations are required to optimise the guided photon yield and to take advantage of the more strongly coupled geometries, such as the 40nm-wide nano trench discussed in section II. With these it should be possible to reach still higher photon yields into the waveguide.

## IV. PHASE SHIFT OF A GUIDED PHOTON BY MOLECULE

When a guided ZPL photon passes a molecule it produces a Stark shift $U(t) = \frac{\hbar g^2 \delta}{\delta^2 + (\Gamma/2)^2 + 2g^2}$, where $g$ is the coupling between the molecule and the field, $\delta$ is the detuning of the light from the molecular resonance, and $\Gamma$ is the total decay rate of the excited molecule. Assuming that the photon is produced by another molecule, as described above, the coupling takes the form $g^2 = g_0^2 e^{-\Gamma t}$. The peak coupling is given by $g_0^2 = \frac{\eta d_x^2 \omega}{2 \hbar n^2 \varepsilon_0 A_{eff} v_g/\Gamma} \left(\frac{n^2+2}{3}\right)^2 = \eta \Gamma_{wg} \Gamma$, where $\eta$ is the branching ratio for the ZPL and the last step makes use of Eq (3). The phase shift induced in the molecule by the passage of this photon is $\frac{1}{\hbar}\int U(t)dt$ and there is an equal and opposite phase shift of the photon itself. If $m$ photons pass the molecule together, we must replace $g_0^2$ by $mg_0^2$. Putting these together we obtain the phase shift per photon due to $m$ photons simultaneously passing one molecule:

$$\varphi(m) = -\frac{\delta}{2 m \Gamma}\ln\left(1 + \frac{2 m \eta \Gamma_{wg} \Gamma}{\delta^2 + (\Gamma/2)^2}\right). \quad (6)$$

For the 785 nm transition of DBT in a matrix of MMA, the spontaneous emission rate at a temperature of 2 K is typically $\Gamma = 2\pi \times 30 \times 10^6$ s$^{-1}$ and the branching ratio for the ZPL line is $\eta = 0.5$ in favourable cases. Taking $\Gamma_{wg} = 0.5 \Gamma$, the phase shift $\varphi(1)$ (dashed) of a single photon passing the molecule peaks at 180 mrad (10°), as shown in Fig. (4), while $\varphi(2)$ (dotted) is appreciably less because of the saturation of the molecule. The difference $\varphi(1) - \varphi(2)$ (solid) peaks at 40 mrad (2°). The real part of the phase shift necessarily brings with it an imaginary part that attenuates the light, reducing the intensity to $exp\left(-\frac{\Gamma}{\delta}\varphi\right)$. The percentage of power lost is also shown in Fig (4) (dash-dotted), where we see that the peak phase shifts are in the region of $20-30\%$ extinction. In principle, this level of nonlinearity is sufficient to provide a useful photon-photon interaction for optical quantum information processing.[32] The differential phase shift increases with stronger coupling arising from reduced mode area. For example, with $0.1 \lambda^2$ mode area using a slot, the maximum differential phase shift rises to almost 8°. Even smaller mode areas may be achieved if plasmonic structures are used. Alternatively, the simple arrangement of a molecule close to the waveguide can be greatly enhanced by forming a waveguide cavity using Bragg mirrors on either side of the molecule. The phase shift is enhanced by the number of bounces of the light, so even a modest cavity is able to produce a large effective photon coupling.

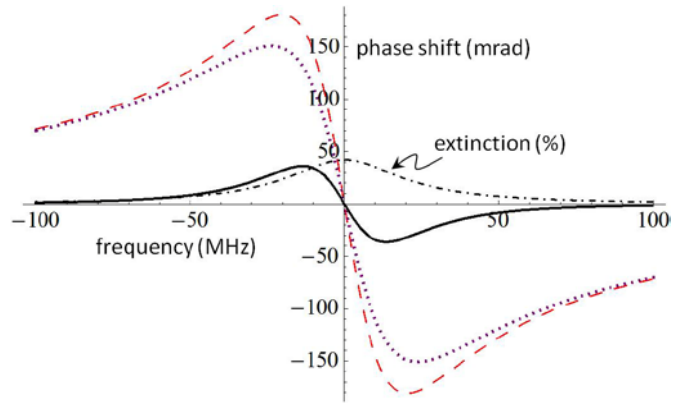

**Fig. 4.** Phase shift and extinction of light in the waveguide due to the presence of one DBT molecule. Dashed line: phase shift $\varphi(1)$ of one photon. Dotted: phase shift per photon $\varphi(2)$ of two photons. Solid line: differential shift $\varphi(1) - \varphi(2)$. Dash-dotted line: extinction

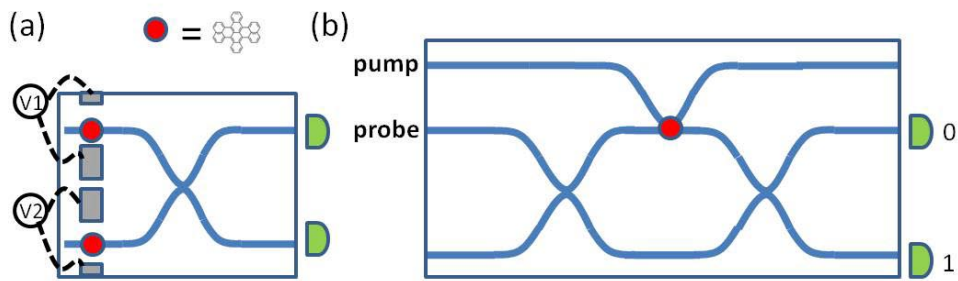

**Fig. 5.** (a) HOM experiment with molecules acting as two single photon sources. Blue lines: waveguides, red dots: single molecules, green boxes: detectors, grey boxes: Stark electrodes, V1, V2: voltage sources. (b) Nonlinear phase gate. Light in the pump input phase shifts light in the upper arm of the interferometer, changing the output port reached by the probe light.

## V. CONCLUSIONS

We have shown that organic molecules placed close to a microfabricated optical strip waveguide offer a very promising addition to optical quantum information processing. In one application, they can act as photon sources integrated on a chip. Figure 5(a) shows a two-photon source, using molecules embedded in Bragg mirrors, coupled to a beam splitter formed by an optical directional coupler.[33] When the molecules are Stark tuned to the same frequency, the identical photons will exhibit Hong-Ou-Mandel interference,[6] and will go as a pair to one or the other detector. The probability of producing a ZPL photon on any given shot of the source is limited by the branching ratio to 50%, however, it may be possible by more subtle engineering of the photonic environment to inhibit emission on the Raman sidebands, or there may be a host-guest system with a more favourable branching ratio.

We have also shown that the molecule can impose a phase shift on each photon, and that this shift can change according to the number of photons because of the saturation of the molecule. Figure 5(b) illustrates a rudimentary gate based on the phase shift. A single photon is fed into a Mach-Zehnder interferometer through the port marked "probe". The probability of arriving at detector 0 (or 1) is determined by the difference in propagation phase for the upper and lower arms. Since a separate "pump" single photon pulse controls the phase shift on the upper arm by saturating the polarisability of the molecule (red dot), there is an effective photon-photon interaction that controls which detector the probe photons reach. It seems on theoretical grounds, associated with the fluctuation-dissipation theorem, that the single photon phase gate may not work as well as described here.[34] However, it may be that in the non-linear response regime of strong coupling this can be circumvented. We see this as a promising area for further research.

## V. ACKNOWLEDGEMENTS

We thank Benoit Bertrand for assistance in the numerical simulations, and Alex Crosse and Myungshik Kim for fruitful discussions.